# Plasmonic resonances and hot spots in Ag octopods

*Ivan I. Naumov\*, Zhiyong Li, and Alexander M. Bratkovsky[#]*

.

Hewlett-Packard Laboratories, 1501 Page Mill Road, Palo Alto, California 94304

\*ivan.naumov@hp.com

[#]alex.bratkovski@hp.com

New type of plasmonic nanoparticles - silver octopods that can be synthesized with a variety of shapes - have been demonstrated to show versatile optical response using the discrete dipole approximation. The octopods show a complex behavior at optical (visible, IR) wavelengths, with three major resonances that can be tuned up to a desired response that makes them especially attractive to use in e.g. high-performance surface enhanced Raman (SERS) detectors. The excited resonant modes strongly depend on the geometrical parameters of the stars, while dependence on their orientation with respect to an incident radiation is moderate, owing to cubic symmetry. The field "hot spots" are formed with the local field enhancement up to 50 times compared to an incident field. They are usually localized at the surface between the arms and may be both "electric" and "magnetic". While the former are of primary importance for SERS, the latter may be identified by trapping magnetic nanoparticles in their vicinity. The results are in very good agreement with the data where available and may be used as a type of a "shape spectroscopy" for the nanoparticles.

In recent years, metallic nanoparticles have been the focus of research due to multitude of possible applications, especially due to possibilities of their surface functionalization, uses as `nanorulers' [1], and due to their supporting plasmonic response and related field amplification by single nanoparticles and their assemblies[2]. The latter is particularly interesting for employing them in the surface enhanced Raman scattering (SERS) studies [3].

Since the surface plasmonic resonances (SPR) are extremely size- and shape-sensitive, the particles with the shapes other than solid sphere seem to be very interesting for further improvement of SERS performance. Due to their high local curvature, presence of sharp points, protrusions, and large aspect ratio, the particles like cubes, disks, nanorods, nanoshells, triangular prisms and multipods can much more effectively amplify the electric field in their vicinity, as compared with spheres[4-8]. Recently, a special attention have been paid to metallic nanostars[9-13]. Very recently, a new type of nanostars, the silver octopods with nearly perfect cubic symmetry have been synthesized[14]. By exposing the initially octahedron-shaped nanoparticles to an etchant with the preferential etching along the [100] direction, the authors managed to obtain *isolated* multi-armed octopod structures maintaining the initial $O_h$ symmetry.

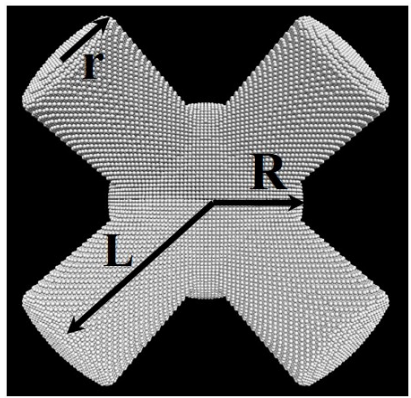

**Fig. 1.** A view of a typical multiarmed silver nanoparticle along the 4-th order symmetry axis, $C_4$. Due to cubic symmetry, the side and top views look similar. $R$, $r$ and $L$ are the geometric parameters defining the structure.

Such a structure can be imagined as a solid sphere with eight protruding cylinders directed along the [111]-type crystallographic directions (Fig. 1).

To understand the plasmonic features of the nanostars, we investigated their scattering properties using the discrete dipole approximation (DDA)[15,16]. In this approach, the object of interest is represented as an array of polarizable spheres on a cubic grid with a lattice period *a*. The period *a* is supposed to be taken much smaller than the wavelength of the incident light ($a \ll \lambda_0$), so that the polarizable spheres could be treated in the quasi-static approximation. Each sphere feels the field of the incident beam $\mathbf{E}_{inc}$ and the fields generated by all other spheres, as described by the following system of linear equations[15,16]:

$$\mathbf{p}_j = \alpha_j \mathbf{E}_j,$$

$$\mathbf{E}_i = \mathbf{E}_{inc,i} + \sum_{j \neq i} \hat{\mathbf{G}}(\mathbf{r}_i - \mathbf{r}_j) \mathbf{p}_j,$$

$$\hat{\mathbf{G}}(\mathbf{R}) \mathbf{p}_j = e^{ikr} \left[ \frac{k^2 (\mathbf{R} \times \mathbf{p}_j) \times \mathbf{R}}{R^3} + (1 - ikr) \frac{3(\mathbf{p}_j \cdot \mathbf{R}) \mathbf{R} - R^2 \mathbf{p}_j}{R^5} \right],$$

where $\mathbf{p}_j$ is the dipole of the sphere centered at the position $\mathbf{r}_j$ with the polarizability $\alpha_j$, $\mathbf{E}_{inc,i} = \mathbf{E}_0 \exp(i\mathbf{k} \cdot \mathbf{r}_i - i\omega t)$ is the incident wave, $\hat{\mathbf{G}}(\mathbf{R} \equiv \mathbf{r}_i - \mathbf{r}_j)$ the tensor Green's function of the Maxwell's equation in free space, $k = \omega/c$. The polarizability is given by the standard Lorenz-Lorentz expression $\alpha_i^{-1} = r_i^{-3}(\varepsilon_i + 2)(\varepsilon_i - 1)^{-1} - 2ik^3/3$ for the sphere with radius $r_i$, with the last complex term giving the radiative correction[17].

The simulated octopods/nanostars have been specified by three geometrical parameters: the core radius *R*, the cylinder radius r and by $L=R+h$, where *h* is the length of the cylinders (Fig. 1). The top area of the cylinders was chosen to be not flat but rounded as a segment of the sphere with the radius *L*. Another possibility to describe the geometry of the octopods is to use two dimensionless parameters *L/R*, *r/R* and an effective radius $a_{eff} = (3V/4\pi)^{1/3}$ expressed via the total volume of the particle *V*. The grid spacing *a* was taken as 2 nm to facilitate convergent results.

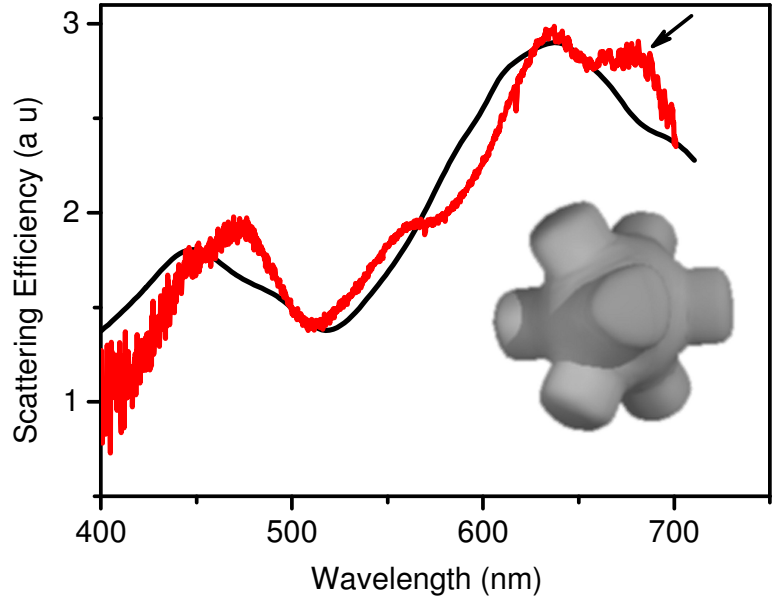

**Fig. 2.** Experimental and theoretical extinction spectra. Shown in red is the experimental curve[14] obtained for a particle in a silica bead (with some uncertainty in geometrical parameters). The theoretical curve (in black) is the optimal fit reached for $a_{eff}$ =70.5 nm, $L/R$=1.4 and $r/R$ =0.6, and the thickness of the bead equal 12 nm. Shown in the insert is an octapod shaped nanostar from Ref. 14. Arrow indicates a shoulder at a longer wavelength that is reproducible with the arms with a triangular crossection.

We begin by demonstrating that the present geometrical model of the nanostars indeed works and is consistent with the data. Shown in red in Fig. 2 is the experimental scattering spectra obtained for a single particle encased in a $SiO_2$ bead[14]. To fit these spectra, we have performed a number of scattering calculations by varying the core size $R$, the arm length $L$, the arm radius $r$, and thickness of the $SiO_2$ cap. The optimal fit was obtained with the following set of geometrical parameters: $a_{eff}$ =70.5 nm, $L/R$=1.4 and $r/R$ =0.6, the silica cap having the thickness of 12 nm (Fig. 2). According to our calculations, the largest linear dimension of the measured nanoparticle (including the shell) was about 200 nm, in a good agreement with an estimation of [14]. Note that the shoulder at longer wavelength than the main peak is due to actual cross section of the arms being close to a *triangular* shape rather than the round one used to generate the theoretical curve in Fig.2.

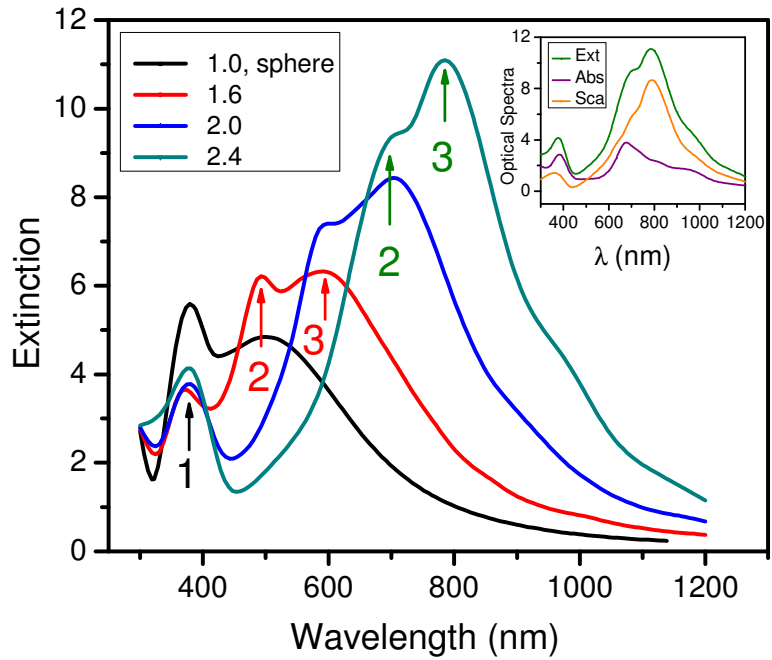

**Fig. 3.** Extinction spectra of sphere and multiarm nanoparticles, all with the same effective radius of 81.4 nm. The ratio of *L/R*, is from left to the right: 1 (sphere), 1.6, 2.0 and 2.4. All the nanostars have the same ratio *r/R* of 0.5. The insert shows how the extinction efficiency is decomposed into absorption and scattering contributions for the case of *L/R*=2.4 as a function of the wavelength λ.

In order to understand how the optical spectra evolve in going from nanosphere to nanostars, we present an extinction efficiency for four nanoparticles having a ratio *L/R* ranging from 1 (a sphere) to 2.4, for the case when the direction of the incident beam is directed along the most symmetrical axis $C_4$, Fig. 3. The size of each nanostar has been chosen so that their volume is equal to that for a 81.4 nm sphere, Fig. 3. The sphere exhibits two well-known peaks[5], the first of which at a wavelength of 375 nm corresponds to a quadrupole plasmonic resonance, whereas the second at 460 nm is dipolar in nature. As the arms just appear, the quadrupole plasmonic resonance starts splitting into two, as it must due to reduction of the initial spherical symmetry $O_3$ to cubic $O_h$. One of these two new peaks (marked as 1) becomes noticeably lower and slightly blue shifts. Another one (indicated as 2), on the contrary, increases in height and moves toward longer wavelengths to a larger extent. The initial dipole plasmon resonance at 460 nm does not split under the reduction of symmetry, but progressively quickly red shifts with *L/R* (the corresponding maximum is marked as 3). This maximum gets progressively higher and at some moment surpasses the height of the peak number two.

Interestingly, the red-most shown quadrupole resonance 2 actually is an *absorption* peak, which does not coincide with the scattering resonance. Similarly, the dipole peak 3 comes, in fact, from the *scattering* component, not from an absorption contribution (similar to the dipole peak for the pure *sphere* in Fig. 3). The fact that the absorption and scattering cross sections do not peak simultaneously means that both the SPRs (2 and 3) are virtual, i.e. correspond to the complex frequencies[18].

Due to high cubic symmetry, the optical properties of the nanostars are expected to be only weakly dependent on the direction of the incident beam **k**. To check this expectation, we compared the extinction spectra for three different **k** vectors directed along the symmetry axes $C_2$, $C_3$, and $C_4$. When averaged over the two mutually-orthogonal directions of polarizations, the extinction spectra (not shown here) exhibit remarkable closeness to each other. This means that *any* beam can excite all the plasmonic modes in the system provided that the light is elliptically or circularly polarized. If the light is linearly polarized, there are only few k directions for which the spectrum is noticeably different for different directions of polarizations. These include the case when **k** is parallel to the second order axis, $C_2$ or [110], with the polarizations along the [1-10] and [001] directions.

Now, we are going to discuss the electromagnetic modes associated with the SPRs 1, 2, and 3 for the case *L/R*=2.4 in Fig. 3 in more detail. Each mode can be characterized by a dipole distribution, which oscillates in time; such distributions corresponding to the initial moment of time are presented in Fig. 4. It is clearly seen that the resonance 1 is a quadrupole in nature, because here approximately half of the electron cloud moves mainly up (red arrows), perpendicular to the axis of arms and half moves mainly down (blue arrows), along the arms. This resonance can be considered as a *bonding hybridization*[11] between the initial sphere quadrupole resonance and the arms' dipolar modes polarized both *parallel* and *perpendicular* to the cylindrical axes. For the *second* plasmon, the resonant currents flow up and down along the arms, but in *opposite* directions for the first four arms (hit by the beam first) compared to the rest four arms (reached by the beam later). Under the symmetry operations of the point group $O_h$, the current field transforms as a row of the three-dimensional irreducible unitary representation $\Gamma'_{25}$ ($T_{2g}$)[19]. Therefore, this is a three-fold degenerate *quadrupole resonance*. And finally, in the SPR 3, the local dipoles oscillate in a similar way in all eight arms. This is, of course, a three-fold degenerate dipole resonance with $\Gamma_{15}$ ($T_{1u}$)[19] symmetry (transforming as vector components)..

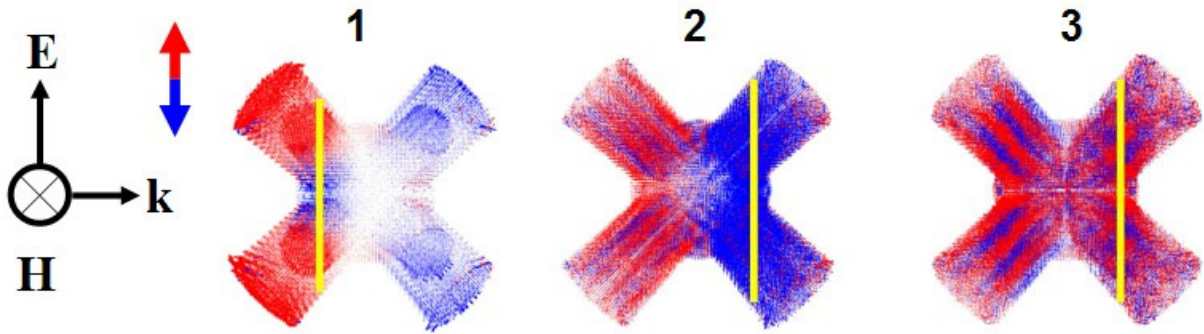

**Fig. 4.** 3D dipole distributions at the initial moment of time for the SPRs 1, 2 and 3. The red and blue arrows show local dipoles directed along or opposite to the external electric field. The hot spots are in the planes marked by the yellow lines.

We should emphasize that in the case of SPR 2 and 3, the more detailed instant dipole distributions in the arms represent a $180^0$ stripe "domain" structure where the "up" and "down" domains run along the arms and alternate along [112] type directions (Fig. 3). With the domain width of approximately 1/3 of the arm radius *r*, such a structure is characterized only by relatively modest polarization charges at the top surface of the cylinders. This is the reason why the strongest enhancement of the electric fields in the nanostars is not a lightning-rod or tip effect (see below).

It is convenient to present hot spots or maxima $|\mathbf{E}|^2$ induced by the resonances on the planes perpendicular to the direction of **k**. The positions of such a plane can be characterized by their dimensionless shortest separation *x* (measured in units of *R*) from the center of the nanostar. We attribute two signs to *x*: *x*<0, if the light beam hits the plane earlier in time than the center, and *x*>0, if later. As analysis shows, the planes contained maximal $|\mathbf{E}|^2$ for the critical frequencies 1, 2, and 3 are defined by *x*=–0.80, 0.85, 0.80, respectively. They are marked by the yellow lines in Fig. 4.

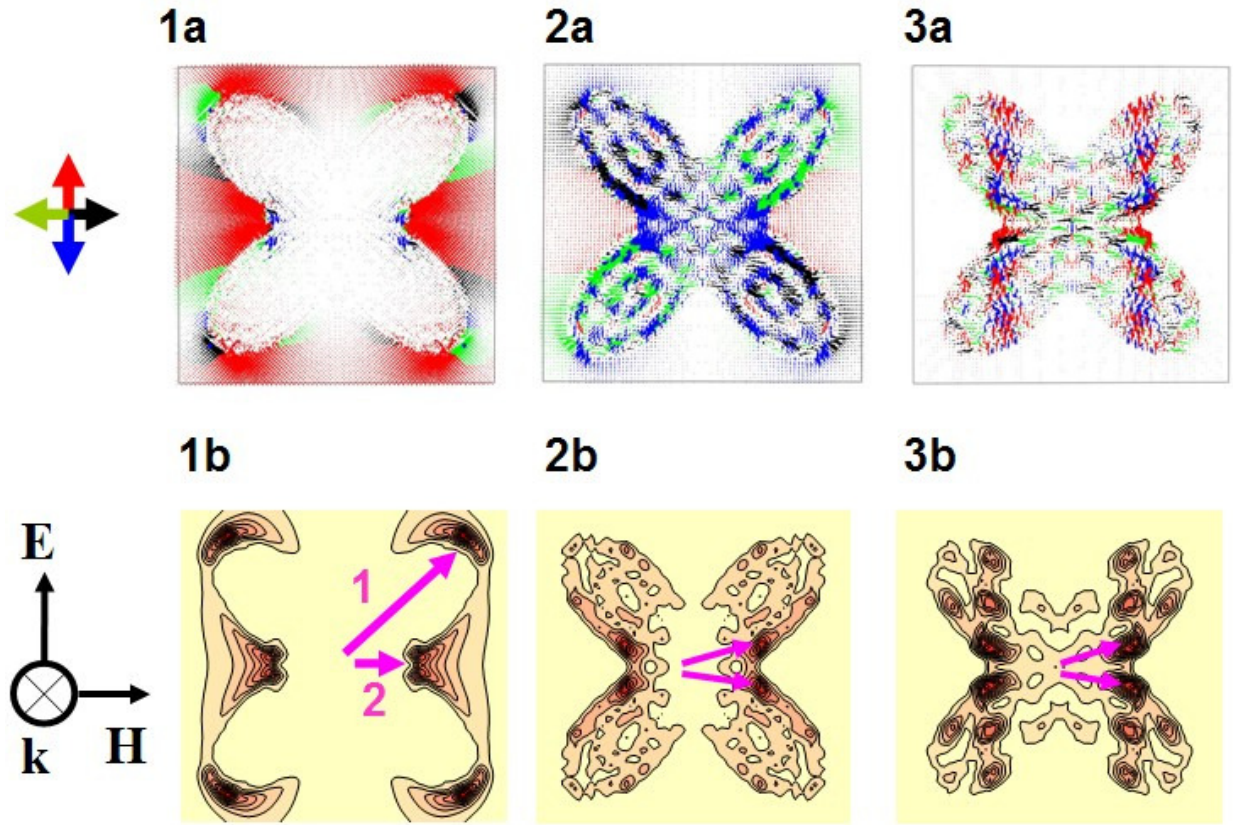

**Fig. 5.** Electric field distributions at the initial moment of time on the planes perpendicular to **k** and containing hot spots (marked by the yellow lines in Fig. 4). The numbers 1, 2, and 3 in the captions correspond to the SPRs in Fig. 3. (a) Vector fields **E** (**r**); shown here in red, blue, green, and black are the vectors **E** directed predominantly up, down, to the left and to the right, respectively. The magnitude of the arrows is proportional to |**E**|. (b) The intensities |**E**|$^2$ averaged over time; the pink arrows point to the hot spots.

For the SPR 1, the field is weak inside the particle and is localized on the surface leading to eight hot spots of two different groups (4 spots in each), Fig. 5. The spots of first group are located on the arms not far from their ends and associated with the local dipolar charge distributions directed *perpendicular* to the arms (panels 1a and 1b). The spots of second group reside between the arms near their confluence; they are induced by the electric current flowing between the upper and lower arms. The near-field enhancement factor associated with the both groups of spots is comparatively modest (about 10, in terms of |**E**|/|**E**$_0$|).

In the case of SPR 2, the field maxima are obtained near the junctions of the arms located *transversely* to the polarization of the incident light (2a and 2b). At these points, the field is directed tangentially to the surface and opposite to the incident field producing the enhancement factor of $|\mathbf{E}|/|\mathbf{E}_0|$ on the order of 40. Accidentally, the positions of hot spots associated with the dipolar resonance 3 (panels 3a and 3b) are very close to those of the resonance 2. However, in passing from the SPR 2 to 3 the character of the electric field changes drastically. Now, in the hot spot region, the field tends to be *parallel* to the polarization of the incident light, while still tangential to the surface. Besides, it becomes more complicated in the core region showing an involved pattern. This resonance produces the *largest* enhancement factor of the field amplitude of about 50.

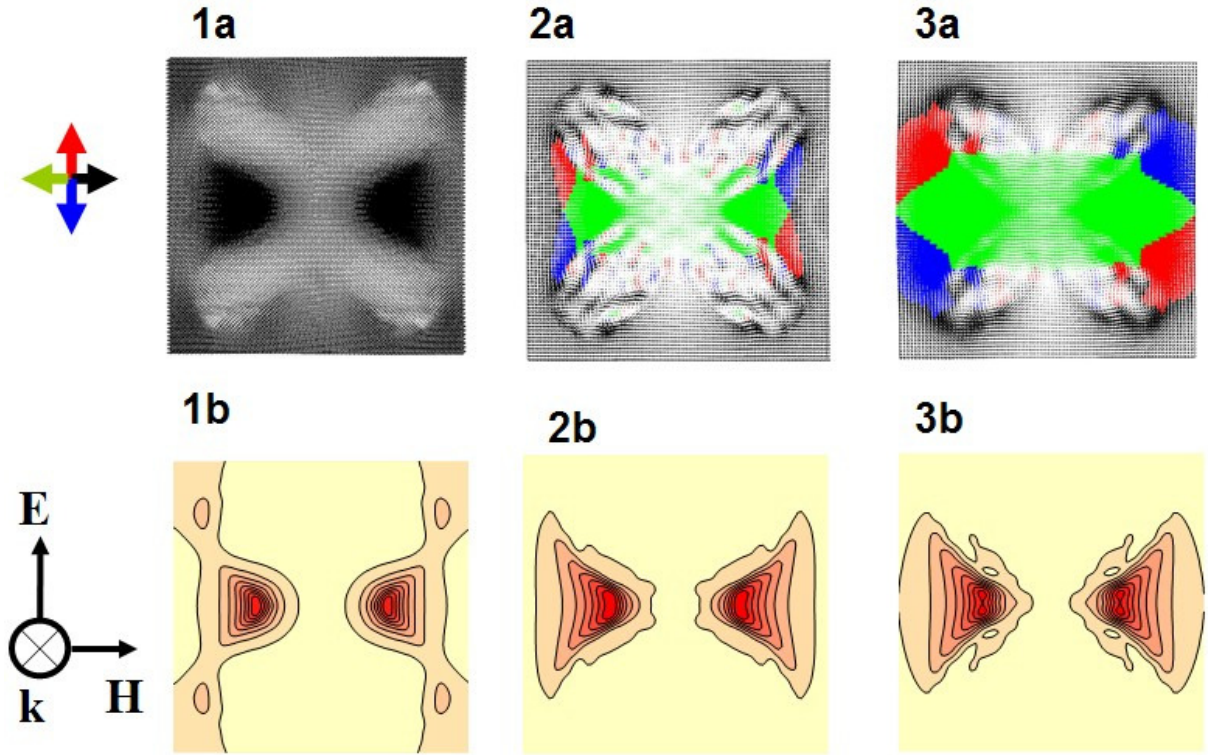

**Fig. 6.** Same as in Fig. 5, but for the magnetic field map.

To understand the origin of electric hot spots, it is necessary to consider the distribution of magnetic fields on the same *x*-planes containing electric hot spots (Fig. 6). As is seen from the figure, the maxima in $|\mathbf{H}|^2$ practically coincide with those in $|\mathbf{E}|^2$, with the only exception of SPR 1, where the spots of the first group do not have their magnetic analogs (compare the panels 1). This means that the hot spots of the second group along with all the other spots (corresponding to the SPRs 2 and 3) are actually magneto-*inductive* in nature. In other words, they stem from the oscillating virtual current loops where the current inside the particle (moving between the lower and upper arms) is shunted by the fields outside of the particle. This conclusion is not surprising, because the linear size of the particles is 5-7 times smaller than the wavelength inside them (for the wavelengths in vacuum λ in the range of 600-800 nm). But in such a case, as pointed by Landau and Lifshitz[20], the scattering properties of the particles are not defined by the excited *electrical* multipoles alone, and the induced electric currents (*magnetic* dipoles) become equally important. It is interesting that in the region of magnetic hot spots, the field **H(r)** is directed along (SPR 1) or against (SPRs 2 and 3) that of incoming light. The latter would correspond of an effective negative permeability at optical frequencies (analogous to a negative permeability response.) Besides, it may be possible to directly visualize those hot *magnetic spots* by trapping small magnetic particles in their vicinity.

It is remarkable that for relatively large *L/R*, the SPRs 2 and 3 are separated practically by the same distance (~100 nm). This trend tends to be preserved even if the geometry of the arms is changed, although the relative strength of the plasmons is strongly sensitive both to the arm length and to its radius. Thus, with increasing *r* the peak 2 in the extinction curve becomes higher and sharper, while the peak 3 is getting lower. By playing only with *L* and *r,* one can form a curve with a two-peak structure in the wide region of wave lengths between 650 and 900 nm. As we discuss below, this tunability is important from the point of view of applications to SERS.

The optical properties of silver nanostars with cubic symmetry offer new interesting possibilities in comparison with the particles of other shapes. First, due to their high symmetry, their plasmonic modes can be excited practically by any light beam, independently of its polarization and direction of propagation. Therefore, the mutual orientation of the target and light becomes much less important than in the case of non-symmetric stars. Second, it is natural to expect that when a molecule is deposited on a nanostar, it may stick somewhere between the arms near their junction. This means that the molecule will find a favorable local environment (hot spots) characterized by strongly enhanced electric fields. And third, as it is well known[21], the Raman scattering enhancement factor scales roughly as $|\mathbf{E}(\omega)|^2 |\mathbf{E}(\omega \pm \Omega)|^2$, where $\Omega$ is the vibrational frequency of the molecule. So, it is important to have a large electric field not only at excited frequency $\omega$, but also at Raman shifted scattering frequencies $\omega_R = \omega \pm \Omega$. The two resonances with close frequencies described above can provide such a realization. In this context, it is very important that the hot spots induced by these resonances are not separated in space but practically coincide.

We should acknowledge many fruitful discussions with Marty Mulvihill, Joel Henzie, Peidong Yang, and our colleagues at IQSL/HP Labs as well as partial support by DARPA.